\documentclass[journal]{IEEEtran}
\usepackage{bm} 
\usepackage{times}  
\usepackage{helvet}  
\usepackage{courier}  
\usepackage[hyphens]{url}  
\usepackage{graphicx} 
\usepackage{pdfpages}
\usepackage{enumitem}
\usepackage{algorithm}
\usepackage{algpseudocode}

\usepackage{diagbox}

\usepackage{amsmath}
\usepackage{newfloat}
\usepackage{listings}
\usepackage{amsfonts}

\usepackage{amsthm}
\usepackage[english]{babel}
\newtheorem{theorem}{Theorem}
\newtheorem{lemma}[theorem]{Lemma}

\usepackage{hhline}
\usepackage{subcaption}
\usepackage{booktabs}

\ifCLASSINFOpdf
\else
\fi
\usepackage{array}
\begin{document}
%

%
\title{Value Functions Factorization with Latent State Information Sharing in Decentralized  Multi-Agent Policy Gradients}
%
%
%
\author{Hanhan~Zhou,
        Tian~Lan,
        and~Vaneet~Aggarwal
\thanks{Hanhan~Zhou and Tian Lan are with the Department of Electrical and Computer Engineering, the George Washington University, Washington, DC, 20052, e-mail: \{hanhan, tlan\}@gwu.edu. Vaneet Aggarwal is with the School of Industrial Engineering and the School of Electrical and Computer Engineering, Purdue University, West Lafayette IN, 47907, email: vaneet@purdue.edu, he is also with the CS Department at KAUST, Saudi Arabia. This research is supported in part by CISCO and Meta.\\  \ \ \ \  © 2023 IEEE.  Personal use of this material is permitted.  Permission from IEEE must be obtained for all other uses, in any current or future media, including reprinting/republishing this material for advertising or promotional purposes, creating new collective works, for resale or redistribution to servers or lists, or reuse of any copyrighted component of this work in other works.}
}
\maketitle

\begin{abstract}
The use of centralized training and decentralized execution for value function factorization demonstrates the potential for addressing cooperative multi-agent reinforcement tasks. QMIX, one of the methods in this field, has emerged as the leading approach and showed superior performance on the StarCraft II micromanagement benchmark. Nonetheless, its monotonic mixing method of combining per-agent estimates in QMIX has limitations in representing joint action Q-values and may not provide enough global state information for accurately estimating single agent value function, which can lead to suboptimal results.
To this end, we present LSF-SAC, a novel framework that features a variational inference-based information-sharing mechanism as extra state information to assist individual agents in the value function factorization. 
We demonstrate that such latent individual state information sharing can significantly expand the power of value function factorization, while fully decentralized execution can still be maintained in LSF-SAC through a soft-actor-critic design. 
We evaluate LSF-SAC on the StarCraft II micromanagement challenge and demonstrate that it outperforms several state-of-the-art methods in challenging collaborative tasks.
We further set extensive ablation studies for locating the key factors accounting for its performance improvements. We believe that this new insight can lead to new local value estimation methods and variational deep learning algorithms.
A demo video and code of implementation can be found at https://sites.google.com/view/sacmm.
\end{abstract}


%
\IEEEpeerreviewmaketitle
\section{Introduction}

\IEEEPARstart{R}EINFORCEMENT learning has been shown to match or surpass human performance in multiple domains, including various Atari games \cite{mnih2015human,duan2016benchmarking,mnih2013playing}, Go \cite{lillicrap2015continuous}, and StarCraft II \cite{vinyals2019grandmaster}. Many real-world problems, like autonomous vehicles coordination \cite{hu2019interaction,he2020data, he2023data,ma2022traffic} and network packet delivery \cite{ye2015multi, miao2021data,luo2022multisource,ma2020statistical} often involve multiple agents’ decision making, which can be modeled as multi-agent reinforcement learning (MARL). Even though multi-agent cooperative problems could be solved by single-agent algorithms, joint state, and action space imply limited scalability\cite{panait2005cooperative,lowe2017multi}. Further, partial observability and communication constraints give rise to additional challenges to MARL problems.
One approach to deal with such issues is the paradigm of centralized training and decentralized execution (CTDE) \cite{kraemer2016multi}. The approaches for CTDE mainly include value function decomposition \cite{sunehag2017value, rashid2018qmix} and multi-agent policy gradient \cite{foerster2018counterfactual}. 

Value decomposition based approaches like QMIX \cite{rashid2018qmix} represent the joint action values using a monotonic mixing function of per-agent estimates. The algorithms recorded the best performance on many StarCraft II micromanagement challenge maps \cite{mahajan2019maven}. Further, it is demonstrated \cite{papoudakis2020comparative} that multi-agent policy gradient is substantially outperformed by QMIX on both multi-agent particle world environment (MPE) \cite{mordatch2018emergence} and StarCraft multi-agent challenge (SMAC) \cite{samvelyan2019starcraft}. Despite recent attempts for combining policy gradient methods and value decomposition, e.g., VDAC \cite{su2021value}, and mSAC \cite{pu2021decomposed}, the achieved improvements over QMIX are limited. One of the fundamental challenges is that the restricted function class permitted by QMIX limits the joint action Q-values it can represent, leading to suboptimal value approximations and inefficient explorations \cite{mahajan2019maven}. A number of proposals have been made to refine the value function factorization of QMIX, e.g., QTRAN~\cite{son2019qtran} and weighted QMIX~\cite{rashid2020weighted}. However, solving tasks that require significant coordination remains as a key challenge.

To this end, we propose {\em LSF-SAC} a Latent State information sharing assisted value function factorization under multi-agent Soft-Actor-Critic paradigm. In particular, we introduce a novel peer-assisted information-sharing mechanism to enable effective value function factorization by sharing the latent individual states, which can be considered extra state information for more accurate individual Q-value estimation by each agent. While global information sharing or communications in MARL - e.g., TarMAC \cite{das2019tarmac}  - typically prevents fully distributed decision making, we show that by leveraging the design of soft-actor-critic, LSF-SAC is able to retain fully decentralized execution while enjoying the benefits of latent individual states sharing. It also incorporates the entropy measure of the policy into the reward to encourage exploration.

The key insight of LSF-SAC is that existing approaches of value function factorization  mainly use the joint state information only in the mixing network, which yet is restricted by the function class it can represent. We show an accurate independent value function estimation requires not only the state information of one specific agent but also a proper represent of all individual state information. We propose a way to extract and utilize the extra state information for individual, per-agent value function estimation through a variational inference method, serving as latent individual state information, since it's impossible and unnecessary to feed the whole state information to individual value function estimations. It is shown to significantly improve the power of value function factorization. Since we utilize such latent state information sharing only in centralized critic, the CTDE assumptions are preserved without affecting fully decentralized decision making, unlike previous work introducing global communications~\cite{wang2019learning}. 
Further, we note that combining actor-critic framework with value decomposition in LSF-SAC offers a way to decouple the decision making of individual agents (through separate policy networks) from value function networks, while also allowing the maximization of entropy to enhance its stability and exploration.

Our key contributions are summarized as follows:
\begin{itemize}[leftmargin=*]
  \item Our novel approach, LSF-SAC, introduces a unique method for value function factorization that incorporates additional individual latent state information to enhance per-agent value function estimation. Our study demonstrates that the inclusion of latent state information can substantially enhance the efficacy of monotonic factorization operators, representing the first framework for value function factorization to leverage this technique.

  \item The soft-actor-critic design in LSF-SAC enables the segregation of policy networks and value function networks for individual agents, allowing a completely decentralized execution while still maintaining the advantages of peer-assisted value function factorization. Additionally, LSF-SAC promotes an entropy maximization approach for multi-agent reinforcement learning, resulting in a more effective exploration. 

  \item Our results showcase the efficacy of LSF-SAC and highlight its superior performance compared to several state-of-the-art baselines on the StarCraft II micromanagement challenge, by achieving better outcomes and faster convergence. 

\end{itemize}

\section{Background}
\subsection{Value Function Decomposition}
Value function decomposition methods \cite{sunehag2017value,rashid2018qmix, son2019qtran,wang2020off} learn  a joint $\mathrm{Q}$ functions $Q^{\operatorname{tot}}(\tau, \mathbf{a})$ as a function of combined individual Q functions, conditioning individual local observation history,then these local $\mathrm{Q}$ values are combined with a learnable mixing neural network to produce joint $Q$ values \cite{shao2021credit}.
\begin{equation}
Q^{\operatorname{tot}}(\tau, \mathbf{a})=q^{\operatorname{mix}}\left(\boldsymbol{s},\left[q^{i}\left(\tau^{i}, a^{i}\right)\right]\right)
\end{equation}

Under the principle of guaranteed consistency between global optimal joint actions and local optimal actions, a global argmax performed on $Q^{\text {tot }}$ yields the same result as a set of individual argmax operations performed on each local $q^{i}$, also known as Individual Global Maximum (IGM):
\begin{equation}
\begin{aligned}
\quad \quad \arg\max _{\boldsymbol{u}} Q^{\operatorname{tot}}=\left(\arg \max _{u_{1}} Q_{1}, \cdots, \arg \max _{u_{N}} Q_{N}\right)
\end{aligned}
\end{equation}
{{VDN \cite{sunehag2017value} takes the joint value function as a summation of local action-value:}
\begin{equation}
Q^{\operatorname{tot}}(\boldsymbol{\tau}, \boldsymbol{u})=\sum_{i=1}^{N} Q_{i}(\tau_{i}, u_{i})
\end{equation}
}
while 
QMIX proposed a more general case of VDN by approximating a broader class of monotonic functions to represent joint action-value functions rather than a summation of the local action values.
\begin{equation}
\frac{\partial Q^{\operatorname{tot}}(\boldsymbol{\tau}, \boldsymbol{u})}{\partial Q_{i}\left(\tau_{i}, u_{i}\right)}>0, \forall i \in \mathcal{N} .
\end{equation}
QPLEX \cite{wang2020qplex} provides IGM consistency by taking advantage of the duplex dueling architecture, 
\begin{equation}
\begin{aligned}
Q^{\operatorname{tot}}(\boldsymbol{\tau}, \boldsymbol{u})=\sum_{i=1}^{N} Q_{i}\left(\boldsymbol{\tau}, u_{i}\right)+\sum_{i=1}^{N}\left(\lambda_{i}(\boldsymbol{\tau}, \boldsymbol{u})-1\right) A_{i}\left(\boldsymbol{\tau}, u_{i}\right)
\end{aligned}
\end{equation}
where 
\begin{equation}
\begin{aligned}
 A_{i}\left(\boldsymbol{\tau}, u_{i}\right) &= w_{i}(\boldsymbol{\tau})\left[Q_{i}\left(\tau_{i}, u_{i}\right)-V_{i}\left(\tau_{i}\right)\right], V_{i}\left(\tau_{i}\right)\\&=
\max _{u_{i}} Q_{i}\left(\tau_{i}, u_{i}\right),  
\end{aligned}
\end{equation}
$w_{i}(\boldsymbol{\tau})$ is a positive weight, yet its operator still limits it to only discrete action space \cite{zhang2021fop}.

\subsection{Maximum Entropy Deep Reinforcement Learning}
In a maximum entropy reinforcement learning framework, also known as soft-actor-critic \cite{haarnoja2018soft}, the objective is to maximize not only the cumulative expected total reward but also the expected entropy of the policy:
\begin{equation}
J(\pi)=\sum_{t=0}^{T} \mathbf{E}_{\left(\mathbf{s}_{t}, \mathbf{a}_{t}\right) \sim \rho_{\pi}}\left[r\left(\mathbf{s}_{t}, \mathbf{a}_{t}\right)+\alpha \mathcal{H}\left(\pi\left(\cdot | \mathbf{s}_{t}\right)\right)\right]
\end{equation}
where  $\rho_{\pi}\left(\mathbf{s}_{t}, \mathbf{a}_{t}\right)$ denotes the state-action marginal distribution of the trajectory induced by the policy $\pi\left(\mathbf{a}_{t}\!|\!\mathbf{s}_{t}\right)$. 
Soft actor-critic utilized actor-critic architecture with independent policy and value networks and an off-policy paradigm for efficient data collection and entropy maximization for effective exploration. It is considered as a state-of-the-art baseline for many RL problems with continuous actions due to its stability and capability.

\subsection{Multi-agent Policy Gradient method}
Multi-agent policy gradient (MAPG) methods are extensions to policy gradient algorithms, with policy $\pi_{\theta_{a}}\left(u^{a} | o^{a}\right)$. Compared with single-agent policy gradient methods, MAPG usually faces the issues of high variance gradient estimates \cite{lowe2017multi} and credit assignment \cite{foerster2017stabilising}. A general multi-agent policy gradient can be written as:
$$
\nabla_{\theta} J=\mathbb{E}_{\pi}\left[\sum_{u} \nabla_{\theta} \log \pi_{\theta}\left(u^{a} | o^{a}\right) Q_{\pi}(s, \mathbf{u})\right]
$$

Current literature on multi-agent policy gradients often leverages centralized training with a decentralized execution (CTDE) approach. This involves using a central critic to obtain additional state information $s$, and helps avoid the high variance associated with vanilla multi-agent policy gradients. For instance, \cite{lowe2017multi} utilize a central critic to estimate $Q\left(s,\left(a_{1}, \cdots, a_{n}\right)\right)$ and optimize parameters in actors by following a multi-agent DDPG gradient, which is derived from:
$$
 \nabla_{\theta_{\alpha}} J=\mathbb{E}_{\pi}\left[\left.\nabla_{\theta_{a}} \pi\left(u^{a} | o^{a}\right) \nabla_{u} \cdot Q_{u^{a}}(s, \mathbf{u})\right|_{u^{\alpha}=\pi\left(o^{\alpha}\right)}\right]
 $$
COMA \cite{foerster2018counterfactual} proposes to apply the following counterfactual policy gradients to solve the credit assignment issue by as:  where $A^{a}(s, \mathbf{u})$ = $\sum_{u^{-}} \pi_{\theta}\left(u^{a} | \tau^{a}\right) Q_{\pi}^{a}\left(s,\left(\mathbf{u}^{-a}, u^{a}\right)\right)$ is the counterfactual advantage for agent $a$. 
\subsection{Variational Autoencoders }
For variables $X\in\mathcal{X}$ which are generated from unknown random variable $z$ based on a generative distribution $p_{u}(x| z)$ with unknown parameter $\boldsymbol{u}$ and a prior distribution on the latent variables, of which we assume is a Gaussian with 0 mean and unit variance $p(z)=\mathcal{N}(\boldsymbol{z}; \mathbf{0}, \boldsymbol{I})$. To approximate the true posterior $p(z | x)$ with a variational distribution $q_{w}(z | \vec{x})=\mathcal{N}(\boldsymbol{z} ; \boldsymbol{\mu}, \boldsymbol{\Sigma}, \boldsymbol{w}) .$ \cite{kingma2014semi,kingma2013auto,papoudakis2020variational} proposed Variational Autoencoders (VAE) to learn this distribution by using the Kullback-Leibler (KL) divergence from the approximate to the true posterior $D_{\mathrm{KL}}\left(q_{w}(z | x) \| p(z | x)\right)$, the lower bound on the evidence $\log p(\boldsymbol{x})$ is derived as:  $\log p(\boldsymbol{x}) \geq \mathbb{E}_{\mathbf{z} \sim q_{w}(z | x)}\left[\log p_{u}(x | z)\right]-D_{K L}\left(q_{w}(z | x) \| p(z)\right)$. \cite{higgins2016beta} proposed $\beta$-VAE, where a parameter $\beta \geq 0$ is used to control the trade-off between the reconstruction loss and the KL-divergence.

\subsection{Information bottleneck Method }

Information bottleneck method \cite{tishby2000information} is a technique in information theory which intorduced as the principle of extracting the relevant information with random input variable $X\in\mathcal{X}$ and output random variable  $Y\in\mathcal{Y}$, while finding the proper tradeoff between extraction accuracy and complexity. Given the joint distribution $p(x,y)$, their relevant information is defined as their mutual information $I(X;Y)$. This problem can also be seen as a rate-distortion problem \cite{tishby2015deep} with non-fixed distortion measure conditioning the optimal map, defined as $$d_{IB} = D_{\mathrm{KL}}(p(y| x) \|p(y| \hat{x}))$$ where $D_{\mathrm{KL}}$ is the Kullback-Leibler divergence. Then the expected IB distortion $E\left[d_{I B}(x, \hat{x})\right] = D_{I B}= I(X ; Y| \hat{X})$, with the variational principle as $$\mathcal{L}[p(\hat{x} | x)]=I(X ; \hat{X}) - \beta I(X ; Y | \hat{X})$$ where $\beta$ is a positive Lagrange multiplier operates as a tradeoff parameter between accuracy and complexity.
\cite{alemi2016deep} further proposed a variational approximation to the information bottleneck using deep neural networks.

\section{Related Works}
Cooperative multi-agent decision-making confronts the situation of exponentially growing joint state and action spaces, which can pose significant challenges \cite{tampuu2017multiagent}. While various strategies such as independent Q-learning and mean field games have been explored in the literature, they often struggle to perform well on complex tasks or require agents with homogenous capabilities \cite{su2021value}. Recently, a centralized training and decentralized execution (CTDE) paradigm has been proposed to tackle these challenges for scalable decision-making \cite{kraemer2016multi}. Key approaches within the CTDE framework include value function decomposition and multi-agent policy gradient methods.

Compared to value-based methods, Policy Gradient methods are generally considered to have more stable convergence and can be extended more easily to continuous action problems \cite{gupta2017cooperative}. One representative approach in the multi-agent Policy Gradient category is COMA \cite{foerster2018counterfactual}, which employs a centralized critic module to estimate an individual agent's counterfactual advantage. However, as highlighted in recent studies \cite{papoudakis2020comparative,son2020qtran++}, value-based methods still outperform multi-agent policy-based methods like MADDPG \cite{lowe2017multi} in the StarCraft multi-agent challenge (SMAC) \cite{samvelyan2019starcraft}.

To address the limitations of centralized critic modules, decomposed actor-critic methods that combine value function decomposition and policy gradient methods with decomposed critics have been introduced to guide policy gradients. VDAC \cite{su2021value} utilizes a structure similar to QMIX to estimate the joint state-value function, while DOP \cite{wang2020off} uses a network similar to Qatten \cite{yang2020qatten} for policy gradients with off-policy tree backup and on-policy TD. However, the authors of \cite{wang2020off} note that decomposed critics are constrained by their limited expressive capability and may not converge to global optima, even if individual policies converge to local optima \cite{zhang2021fop}. Although extensions of the monotonic mixing function, such as QTRAN~\cite{son2019qtran}, and weighted QMIX~\cite{rashid2020weighted}, have been explored, significant challenges remain when tackling tasks that require substantial coordination. 

Another related topic is representational learning for reinforcement learning, and various methods have been proposed to learn effective state representations. For instance, \cite{ha2018world} proposed a VAE-based forward model to learn state representations in the environment. \cite{grover2018learning} developed a technique to learn Gaussian embedding representations of different tasks during meta-testing. \cite{igl2018deep} introduced a recurrent VAE model that encodes observation and action history and learns a variational distribution of the task. Authors in \cite{chen2022relax, chen2022explain,pac} use counterfactual information as explanations of deep reinforcement learning agents. 

As also analyzed and suggested in MAVEN \cite{mahajan2019maven} and QTRAN \cite{son2019qtran}, the representational constraints on the joint action-values introduced by the monotonic mixing network in QMIX \cite{rashid2020weighted} and similar methods will lead to provably poor exploration and sub-optimal behavior policies. To solve this issue, one of directions is to release the restriction of the joint action-value functions, e.g., QTRAN uses a linear summation over the utility functions and an additional value estimation, WQMIX \cite{rashid2020weighted} uses an unrestricted joint action-value function estimator as the weighted projection of a wider class of joint action-value functions, REMIX \cite{mei2023remix} considers a regret minimization method to acquire this optimal weights; another direction is to promote a more committed exploration algorithm to recover the poor exploration introduced by the monotonic constraints, e.g., MAVEN combines value and policy-based methods with agents conditioning their behavior on a variable controlled policy for a temporally extended exploration, MAC-PO \cite{10.5555/3545946.3598672} proposes optimal prioritized experience replay for improved multi-agent tasks. In this work, our proposed Decomposed Soft-actor-critic will promote the exploration through entropy maximization, while providing additional information from latent state information as assisted information for value function factorization.

{{{Another topic related is communication-based MARL methods. Although the requirement of communication abilities might limit the actual use case of the proposed algorithm, with communications enabled, MARL agents will have a better understanding of the environment (or the other agents) and are therefore able to coordinate their behaviors and potentially better performances.
Most works leverage local information to generate an encoded message. The messages may contain individual observations \cite{niu2021multi,du2021learning}, or intended actions (or plans) \cite{jiang2018learning,wang2019learning}
}}}. A close paper to our work is NDQ \cite{wang2019learning}, which also utilizes latent variables to represent the information as the communication messages during the decentralized agent's execution. Although we both consider information extraction as an information bottleneck problem, there are several key differences between our work and NDQ: (I) NDQ is a value-based method, while our work is a policy-based method under the soft-actor-critic framework. (II) NDQ requires communication between agents during decentralized execution, which limits its use cases. At the same time, we only utilize the latent extra state information during the central critics so that CTDE is maintained. (III) NDQ requires one-to-one communication during the execution stage, while in this work, we introduce a latent information-sharing mechanism that can be considered as an all-to-all message-sharing method. {{{By enabling the latent information sharing mechanism in our work as a communication method, this work could potentially be transformed to a communication-based method, and many communication-based methods can be transformed into a framework where communication is only used for centralized training and restricted during execution, nevertheless, their performance and the actual use case may vary a lot.}}}

The proposed LSF-SAC method leverages an actor-critic design with latent state information for value function factorization. We introduce a novel way to utilize the extra state information, as inspired from $\beta$-VAE \cite{higgins2016beta}, by using variational inference in a decomposed critic as latent state information for better individual value estimation. Despite information sharing, CTDE is still maintained due to the use of an actor-critic structure. We also utilize the entropy and expected return maximization for better exploration through soft actor-critic with separate actor and critic networks. 

\section{System Model}
We approach the problem as a fully cooperative multi-agent environment with a decentralized partially observable Markov decision process (DEC-POMDP) \cite{oliehoek2016concise}. The DEC-POMDP is defined as given by a tuple $G=\langle I, S, U, P, r, Z, O, n, \gamma\rangle$, where $I\!\equiv\{1,2, \cdots, n\}$ is the finite set of agents. 
The state of the system is defined as a finite set of global states $s \in S$, from which each agent draws its own observation from the observation function $o_{i} \in O(s,i): S \times A \!\! \rightarrow \!\! O$. At each timestamp $t$, each agent $i$ chooses an action $u_{i} \in U$ where $U$ is a set of actions available, forming a joint action selection  $\boldsymbol{u}$. A shared reward is then given as $r\!\!=\!\!R(s,\boldsymbol{a}) : S \times \!\! \mathbf{U}\!\!\! \rightarrow \!\! \mathbb{R} $, and each agent transitions to a new state $s^{\prime}$ based on the transition probability function $P\left(s^{\prime} | s, \mathbf{u}\right) : S \times U \rightarrow [0,1]$.ach agent maintains its own action-observation history  $\tau_{i} \in \mathrm{T} \equiv(O \times U)^{*}$. Then a joint action value function $Q_{t o t}^{\pi}(\boldsymbol{\tau}, \boldsymbol{u})=\mathbb{E}_{s_{0: \infty}}, \boldsymbol{u}_{0: \infty}\left[\sum_{t=0}^{\infty} \gamma^{t} r_{t} | s_{0}=s, \boldsymbol{u}_{0}=\boldsymbol{u}, \boldsymbol{\pi}\right]$ is proposed with policy $\pi$, and $\gamma \in[0,1)$ is the discount factor. Notation in bold represents joint quantities across all agents, and quantities with superscript $i$ are specific to agent $i$.
 
\begin{figure*}[htbp]
\includegraphics[width=0.96\textwidth]{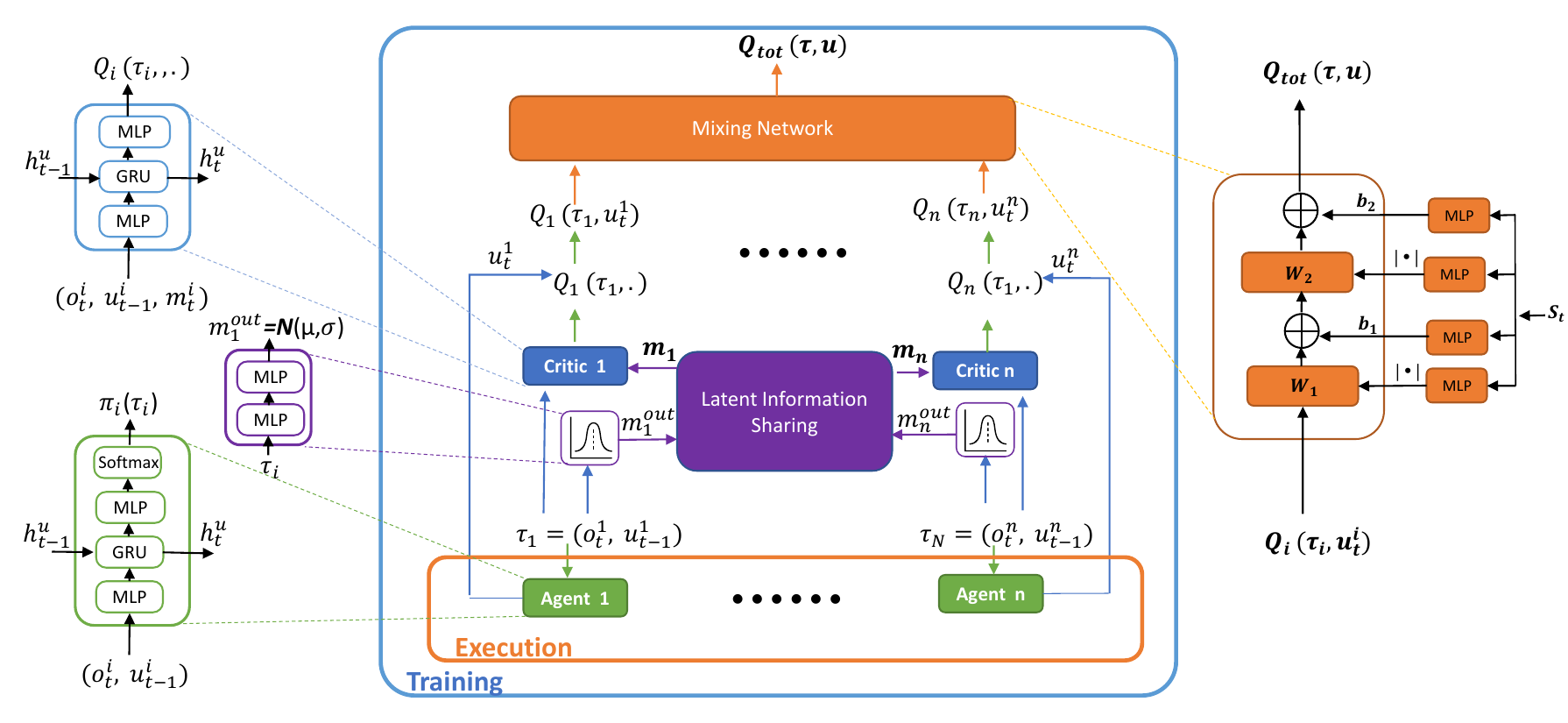}
\caption{Overview of LSF-SAC Approach. Best viewed in color.}
\end{figure*}

\section{Proposed Approach}

In this section, we first introduce the main structure of our proposed method, LSF-SAC, then we discuss the detailed implementation of the key designs, namely soft actor-critic framework for multi-agent reinforcement learning and value decomposition with latent information-sharing mechanism, and their corresponding optimizing strategies. 

\subsection{Framework Overview}
In our learning framework (Fig.~1), each individual actor (Green part) outputs $\pi_{\theta}(a^{i}|\tau ^{i})$ only conditioned on its own local observation history. The centralized mixing network (Orange Part)  approximates the joint action-value function from individual value functions (Blue part). A latent information-sharing mechanism (Purple part) is proposed to encode the extracted extra state information to assist individual agents in local action-value estimation. Function approximators (neural networks) are used for both actor and critic networks and optimized with stochastic gradient descent.

The centralized critic network consists of (i) a local Q-network for each agent, (ii) a mixing network that takes all individual action-values with their weights and biases generated by a separate hyper-network, and (iii) an extra state information encoder to generate latent state information for facilitating individual Q-value estimation. For each agent $i$, the local Q network represents its local Q value function $q_{i}(\tau _{i}, a_{i}, m_{i})$ where $m_{i}$ is the extra state information for agent $i$ drawn from the global information sharing pool. More precisely, the information for agent $i$ is generated from the messages of all other agents following a multivariate Gaussian distribution, denoted as  $m_{i}=<\!\!m_{1}^{out}\cdots  m_{i}^{out} \cdots m_{n}^{out}\!\!>$ with $m_{i}^{out} \sim  \boldsymbol{N}(f_{m}(\tau _{i};\theta _{m}),\boldsymbol{I}))$, where $\tau_{i}$ is the local observation history, $\theta_{m}$ is the parameters of encoder $f_{m}$ and $\boldsymbol{I}$ is an identity matrix. 

The mixing network is a feed-forward network, following the approach in QMIX, which mixes all local Q values to produce an estimate $Q^{tot}$. The weights and biases of the mixing network are generated by a hypernetwork that takes joint state information $s$. To enforce monotonicity, the weights generated from the hyper-networks are followed by an absolute function to create non-negative values. The decentralized actor network is similar to the individual Q network, except it only conditions on its own observation and action history, and a softmax layer is added to the end of the network to convert logits into categorical distribution. 
The overall goal is to minimize: 
\begin{equation}
\mathcal{L}(\boldsymbol{\theta})=\mathcal{L}_{TD}(\boldsymbol{\theta_{TD}})+\lambda_{1} \mathcal{L}_{m}\left(\boldsymbol{\theta}_{m}\right) + \lambda_{2} \mathcal{L}_{\pi }\left(\boldsymbol{\theta}_{\pi }\right)
\end{equation}
where $\mathcal{L}_{TD}(\boldsymbol{\theta_{TD}})$ is the TD loss, of which we show it can also be used as the center critic loss, $\mathcal{L}_{m}\left(\boldsymbol{\theta}_{m}\right)$ is the message encoding loss, and $\mathcal{L}_{\pi }\left(\boldsymbol{\theta}_{\pi }\right)$ is the joint actor (policy) loss. $\lambda_{1}$ and $\lambda_{2}$ are the weighting terms. The details about latent state information generation and soft-actor-critic framework along with how to optimize them will be discussed in the following section.

\subsection{Variational Approach Based Latent State Information}
One of the key advantages of multi-agent policy gradients under the CTDE assumption is the effective utilization of extra state information. In our design, not only is the extra state information accessible to the mixing network but also to the individual agents' value networks (through information sharing). Due to the partial observability and uncertainty of the multi-agent environments, the individual value estimation conditioned on its own observation and action history can be volatile and unreliable. Intuitively, introducing extra information from other agents helps remove the ambiguity and uncertainty of current observation to enable effective individual value estimation.

However, it remains a crucial problem on how to efficiently and effectively encode such extra state information.  {{{In most scenarios, even during the centralized training stage, it is impossible to directly feed the whole state information as input for individual value functions, as it consists of other agents' observation and unseen state information, without a carefully designed algorithm it is hard for a local agent to utilize them; at the same time, the input size of global state information is significantly larger than local observations, which would make the training longer to converge}}}. We consider this as an information bottleneck problem \cite{tishby2000information}, specifically, for agent $i$, we maximize the mutual information between other agents’  encoded information and their actions while minimizing the mutual information between its own encoded information and action selection, so that only the necessary information is chosen and then efficiently encoded. 

To encode additional state information for estimating individual values in an efficient and effective manner, we approach this problem as an information bottleneck problem \cite{tishby2000information}, and the objective for each agent $i$ can be written as:
\begin{equation}
J_{m}\left(\boldsymbol{\theta}_{m }\right)=\sum_{j=1}^{n}\left[I_{{\theta}_{m}}\left(A_{j} ; M_{i} | \mathrm{T}_{j}, M_{j}\right)-\beta I_{\boldsymbol{\theta}_{m}}\left(M_{i};T_{i}\right)\right]
\end{equation}
where $A_{j}$ is agent $j$'s action selection, $M_{i}$ is a random variable of $m_{i}^{out}$, $T_{j}$ is a random variable of $\tau_{j}$, and a parameter $\beta \geq $ 0 is used to control the trade-off between the mutual information of its own and other agents. 
However, since the mutual information is intractable, this does not result in a model that can be learned. To overcome this challenge, we utilize variational approximation techniques, specifically the deep variational information bottleneck approach \cite{alemi2016deep}. By parameterizing our model with a neural network, we can derive and optimize a variational lower bound for the first term of our objective function, as follows. Detailed derivations and proofs can be found in Appendix {\bf A.1}.

\begin{lemma}
A lower bound of mutual information $I_{{\theta}_{m}}(A_{j} ; M_{i} | \mathrm{T}_{j}, M_{j})$ is  
\[
\mathbb{E}_{\mathbf{T} \sim \mathcal{D}, M_{j} \sim f_{m}} [-\mathcal{H}[p(A_{j} | \mathbf{T}), q_{\psi }(A_{j} | \mathrm{T}_{j}, \boldsymbol{M})]]
\]
where $q_{\psi}$ is a variational Gaussian distribution with parameters $\psi$ to approximate the unknown posterior $p(A_{j} | \mathrm{T}_{j}, M_{j})$, $\mathbf{T}=\{T_{1},T_{2}, \cdots, T_{n}\}$, $\mathbf{M}=\{M_{1},M_{2}, \cdots, M_{n}\}$. 
\end{lemma}
\begin{proof} We provide a proof outline as follows. 
\begin{equation*}
\begin{aligned}
& I_{\theta_{c}}\left(A_{j} ; M_{i} | T_{j}, M_{j}\right) \\
=& \int d a_{j} d \tau_{j} d m_{j} p\left(a_{j}, \tau_{j}, m_{j}\right) \log \frac{p\left(a_{j} | \tau_{j}, m_{j}\right)}{p\left(a_{j} | \tau_{j}, m_{j}^{out}\right)} 
\end{aligned}
\end{equation*}
where $p(a_{j}|\tau_{j}, m_{j})$ is fully defined by our decoder $f_{m}$ and Markov Chain\cite{littman1994markov}. Note this is intractable in our case, let $q_{\psi}(a_{j} | \mathrm{\tau}_{j}, m_{j})$ be a variational approximation to $p(a_{j}| \tau_{j}, m_{j})$. Since the KL-divergence is always positive,

hence
$$
\begin{aligned}
& I_{\theta_{c}}(A_{j} ; M_{i} | T_{j}, M_{j}) \\
& \geq \int d a_{j} d \tau_{j} d m_{j} p\left(a_{j}, \tau_{j}, m_{j}\right) \log \frac{q_{\psi}\left(a_{j} | \tau_{j}, m_{j}\right)}{p\left(a_{j} | \tau_{j}, m_{j}^{out}\right)} \\
& = \mathbb{E}_{\mathbf{T} \sim \mathcal{D}, M_{j} \sim f_{m}}
 [-\mathcal{H}[p(A_{j} | \mathbf{T}), q_{\psi }(A_{j} | \mathrm{T}_{j}, \boldsymbol{M})]] \\
& +  \mathcal{H}(A_{j}|T_{j},M_{j}^{out})
\end{aligned}
$$
Consider $\mathcal{H}(A_{j}|T_{j},M_{j}^{out})$ is a positive term that is independent of our optimization procedure and can be ignored, then we have
\begin{equation}
\begin{array}{l}
I_{{\theta}_{m}}\left(A_{j} ; M_{i} | \mathrm{T}_{j}, M_{j}\right)  \\ 
 \quad \geq \mathbb{E}_{\mathbf{T} \sim \mathcal{D}, M_{j} \sim f_{m}}
 \left[-\mathcal{H}\left[p\left(A_{j} | \mathbf{T}\right), q_{\psi }\left(A_{j} | \mathrm{T}_{j}, \boldsymbol{M}\right)\right]\right]
\end{array}
\end{equation}
\end{proof}
Similarly, by introducing another variational approximator $q_{\phi}$, we have  
\begin{equation}
\begin{array}{l}
\begin{aligned}
I_{\boldsymbol{\theta}_{m}}\left(M_{i};T_{i}\right)  
& = \mathbb{E}_{\mathrm{T}_{i} \sim D,M_{j} \sim f_{m}}\left[D_{\mathrm{KL}}\left(p\left(M_{i}|\mathrm{T}_{i}\right) \| p\left(M_{i}\right)\right)\right]
\\
& \leq \mathbb{E}_{\mathrm{T}_{i} \sim D,M_{j} \sim f_{m}}\left[D_{\mathrm{KL}}\left(p\left(M_{i}|\mathrm{T}_{i}\right) \| q_{\phi}\left(M_{i}\right)\right)\right] 
\end{aligned}

\end{array}
\end{equation}
where $D_{\mathrm{KL}}$ denotes the Kullback-Leibler divergence operator and $q_{\phi}( M_{i})$ is a variational posterior estimator of $p(M_{i})$ with parameters $\phi$ (see Appendix {\bf A.1} for details).  Then with the evidence lower bound derived above we optimize this bound for the message encoding objective which is to minimize 

\begin{equation}
\begin{array}{l}
\begin{aligned}
\mathcal{L}_{m}(\boldsymbol{\theta}_{m})&=\mathbb{E}_{\mathbf{T} \sim \mathcal{D}, M_{j} \sim f_{m}}
[-\mathcal{H}[p(A_{j}| \mathbf{T}), q_{\psi }(A_{j} | \mathrm{T}_{j}, M_{j})]\\
&+\beta D_{\mathrm{KL}}(p(M_{i} | \mathrm{T}_{i}) \| q_{\phi}(M_{i}))].
\end{aligned}
\end{array}
\end{equation}

\begin{algorithm}[ht]
    \caption{LSF-SAC  }
    \begin{algorithmic}[1]
        \For{$k = 0 $ to $train\_steps\_limits$}                    
            \State {Reset environment}
            \For{$t = 0 $ to $max\_episode$}    
               \State For each agent $i$, choose action $a_{i}\sim \pi_{i}$
               \State Execute joint action $\mathbf{a}$, record reward $r$,
               
               \Statex \quad\quad\quad save state-action history $\boldsymbol{\tau}$, next state $s_{t+1}$
               \State Store ($\boldsymbol{\tau}$, $\boldsymbol{a}, r, \boldsymbol{\tau^{'}}$) in replay buffer  $\mathcal{D}$
            \EndFor
        \For{t = 1 to T}
            \State Sample minibatch $\mathcal{B}$ from $\mathcal{D}$
            \State Generate latent state information
            \Statex \quad\quad\quad $m_{i}^{out}\!\sim\boldsymbol{N}(f_{m}(\tau _{i};\theta _{m}),\boldsymbol{I}))$, for $i = 0$ to $n$
            \State Update critic network
            \Statex \quad\quad\quad $\boldsymbol{\theta_{TD}} \gets \eta\hat{\nabla}\mathcal{L}_{TD}(\boldsymbol{\theta_{TD}})$ w.r.t Eq(9)
            \State Update policy network 
            \Statex \quad\quad\quad $\boldsymbol{\pi} \gets \eta\hat{\nabla}\mathcal{L}(\boldsymbol{\pi})$ w.r.t Eq(7)

            \State Update encoding network 
            \Statex \quad\quad\quad $\boldsymbol{\theta}_{m} \gets \eta\hat{\nabla}\mathcal{L}_{m}(\boldsymbol{\theta}_{m})$ w.r.t Eq(5)
            \State  Update temperature parameter
            \Statex \quad\quad\quad  $\alpha \gets \eta \hat{\nabla}\alpha$ w.r.t Eq(8)
            \If{$time\_to\_update\_target\_network$}
                \State $\boldsymbol{\theta^{-}} \gets \boldsymbol{\theta} $
            \EndIf
        \EndFor
        \EndFor
        \State Return $\boldsymbol{\pi}$
    \end{algorithmic}
\end{algorithm}

\subsection{Factorizing Multi-Agent Maximum Entropy RL} 
In this section, we present one possible implementation of expanding soft actor-critic to multi-agent domain with latent state information assisted value function decomposition, its objective extended to multi-agent domain can be defined as 
\begin{equation}
J(\pi)=\sum_{t} \mathbb{E}\left[r\left(\mathbf{s}_{t}, \mathbf{a}_{t}\right)+\alpha \mathcal{H}\left(\pi\left(\cdot | \mathbf{s}_{t}\right)\right)\right]
\end{equation}
where the temperature $\alpha$ is the hyper-parameter to control the trade-off between maximizing the expected return and maximizing  the entropy for better exploration. 

Following the previous research on value decomposition, to maximize both the expected return and the entropy, we find the soft policy loss of LSF-SAC as:
\begin{equation}
\begin{array}{l}
\begin{aligned}
\mathcal{L}_{LP}(\pi) &=\mathbb{E}_{\mathcal{D}}\left[\alpha \log \boldsymbol{\pi}\left(\boldsymbol{u}_{t} | \boldsymbol{\tau}_{t}\right)-Q^{\pi}_{tot}\left(\boldsymbol{s_{t}}, \boldsymbol{\tau_{t}}, \boldsymbol{u}_{t}, \boldsymbol{m}_{t}\right)\right] \\
&= -q^{\operatorname{mixing}}\left(\boldsymbol{s}_{t}, \mathbb{E}_{\pi^{i}}\left[q^{i}\left(\tau_{t}^{i}, u_{t}^{i}, m_{t}^{i}\right)-\alpha \log \pi^{i}\left(u_{t}^{i} | \tau_{t}^{i}\right)\right]\right)
\end{aligned}
\end{array}
\end{equation}
where $q^{mixing}$ is the value decomposition operator with $u_{i}\sim \pi_{i}(o_{i})$, and $\mathcal{D}$ is the replay buffer used to sample training data (state-action history and reward, etc.).

Then, we can tune the temperature $\alpha$ as proposed in \cite{haarnoja2018soft} by optimizing the following:
\begin{equation}
J(\alpha) = \mathbb{E}_{a_{t} \sim \pi_{t}} [-\alpha \log \pi_i(a_{t} | s_{t}) - \alpha \mathcal{H}_0]
\end{equation}
Unlike VDAC which shares the same network for actor networks and local Q value estimations, we use a separate network for policy networks and train them independently from critic networks. Latent state information is used for individual critics for joint action value function factorization. We propose a latent state information assisted soft value decomposition design as 

\[
 Q_{tot}(\boldsymbol{\tau}, \boldsymbol{a}, \boldsymbol{m} ; \boldsymbol{\theta}) = q^{\operatorname{mixing}}(\boldsymbol{s}_{t}, \mathbb{E}_{\pi^{i}}[q^{i}(\tau_{t}^{i}, a_{t}^{i},m_{t}^{i}); \boldsymbol{\theta}])
\]

We then use TD advantage with latent information sharing the design as the critic loss, i.e.,
\begin{equation}
\begin{aligned}
\mathcal{L}_{TD}(\boldsymbol{\theta}) &=[r\!+\!\gamma \max _{\boldsymbol{a}^{*}} Q_{tot}\left(\boldsymbol{\tau}^{\prime}, \boldsymbol{a}^{\prime}, \boldsymbol{m}^{\prime} ; \boldsymbol{\theta}^{-}\right)\!-\! Q_{tot}^{\pi}(\boldsymbol{\tau}, \boldsymbol{a}, \boldsymbol{m} ; \boldsymbol{\theta})]^{2}
\\
&= [r\! +\! \gamma \max _{\boldsymbol{a}^{*}} q^{\operatorname{mixing}}(\boldsymbol{s}_{t}, \mathbb{E}_{\pi^{i}}[q^{i}(\tau_{t+1}^{i}, a_{t+1}^{i},m_{t+1}^{i}); \boldsymbol{\theta^{-}}])
\\
& \quad- q^{\operatorname{mixing}}(\boldsymbol{s}_{t}, \mathbb{E}_{\pi^{i}}[q^{i}(\tau_{t}^{i}, a_{t}^{i},m_{t}^{i}); \boldsymbol{\theta}])]^{2}
\end{aligned}
\end{equation}
where  $a_{i}\sim \pi_{i}(o_{i})$, $\boldsymbol{\theta}^{-}\ $ is the parameters of the target network that are periodically updated.  Detailed derivations can be found in  Appendix {\bf A.2}.
\section{Experiments}

In this section, we first empirically study the improvements of power in value function factorization achieved by LSF-SAC through a non-monotonic matrix game. We compare the results with several existing value function factorization methods. Then in StarCraft II, we compare LSF-SAC with several state-of-the-art baselines. Finally, we perform several ablation studies to analyze the factors that contribute to the performance.

\subsection{Single-state Matrix Game}
Proposed in QTRAN~\cite{son2019qtran}, the non-monotonic matrix game, as illustrated in Table~1(a), consists of two agents with three available actions and a shared reward. We show the value function factorization results of QTRAN, LSF-SAC, VDN, QMIX, and DOP~\cite{wang2020off}.

\begin{table}[ht]
\begin{subtable}[c]{0.23\textwidth}
\centering\resizebox{\textwidth}{!}{
\begin{tabular}{|c||c|c|c|}
\hline
\diagbox{$u_{1}$}{$u_{2}$}  & A   & B   & C   \\ \hline\hline
A & \textbf{8.0}   & -12.0 & -12.0 \\ \hline
B & -12.0 & 0.0   & 0.0   \\ \hline
C & -12.0 & 0.0   & 0.0  \\ \hline
\end{tabular}}
\subcaption{Payoff of matrix game}
\end{subtable}
\hfill
\begin{subtable}[c]{0.23\textwidth}
\centering\resizebox{\textwidth}{!}{
\begin{tabular}{|c||c|c|c|}
\hline
\diagbox{$Q_{1}$}{$Q_{2}$} & \textbf{4.2}(A)   & 2.3(B)   & 2.3(C)   \\ \hline\hline
\textbf{3.8}(A)& \textbf{8.0}   & 6.13 & 6.1 \\ \hline
-2.1(B) & 2.1 & 0.2   & 0.2  \\ \hline
-2.3(C) & 1.9 & 0.0   & 0.0   \\ \hline
\end{tabular}}
\subcaption{QTRAN}
\end{subtable}
\hfill
\begin{subtable}[c]{0.23\textwidth}
\centering\resizebox{\textwidth}{!}{
\begin{tabular}{|c||c|c|c|}
\hline
\diagbox{$Q_{1}$}{$Q_{2}$} & \textbf{1.7}(A)   & -11.5(B)   & -12.7(C)   \\ \hline\hline
\textbf{0.4}(A) & \textbf{8.1}   & -6.2 & -6.0 \\ \hline
-9.9(B) & -6.0 & -5.9   & -6.1   \\ \hline
-9.5(C) & -5.9 & -6.0   & -6.0   \\ \hline
\end{tabular}}
\subcaption{LSF-SAC}
\end{subtable}
\hfill
\begin{subtable}[c]{0.23\textwidth}
\centering\resizebox{\textwidth}{!}{
\begin{tabular}{|c||c|c|c|}
\hline
\diagbox{$Q_{1}$}{$Q_{2}$} & 3.1(A)   & \textbf{-2.3}(B)   & -2.4(C)   \\ \hline\hline
-2.3(A) & -5.4   & -4.6 & -4.7 \\ \hline
-1.2(B) & -4.4 & -3.5   & -3.6   \\ \hline
\textbf{-0.7}(C) & -3.9 & \textbf{-3.0}   & -3.1   \\ \hline
\end{tabular}}
\subcaption{VDN}
\end{subtable}
\hfill
\begin{subtable}[c]{0.23\textwidth}
\centering\resizebox{\textwidth}{!}{
\begin{tabular}{|c||c|c|c|}
\hline
\diagbox{$Q_{1}$}{$Q_{2}$} & -0.9(A)   & 0.0(B)   & \textbf{0.0}(C)   \\ \hline\hline
-1.0(A) & -8.1   & -8.1 & -8.1 \\ \hline
\textbf{0.1}(B) & -8.1 & 0.0   & \textbf{0.0}   \\ \hline
0.1(C) & -8.1 & 0.0   & 0.0   \\ \hline
\end{tabular}}
\subcaption{QMIX}
\end{subtable}
\hfill\begin{subtable}[c]{0.23\textwidth}
\centering\resizebox{\textwidth}{!}{
\begin{tabular}{|c||c|c|c|}
\hline
\diagbox{$Q_{1}$}{$Q_{2}$} & -2.5(A)   & -1.3(B)   & \textbf{0.0}(C)   \\ \hline\hline
-1.0(A) & -7.8   & -6.0 & -4.2 \\ \hline
\textbf{0.1}(B) & -6.1 & -4.4   & \textbf{-2.6}   \\ \hline
0.1(C) & -4.2 & -2.4   & -0.7   \\ \hline
\end{tabular}}
\subcaption{DOP}
\end{subtable}
\hfill
\caption{Payoff Matrix of the one-step matrix game, $Q_{1}, Q_{2}$ and reconstructed $Q_{tot}$ of selected algorithms. The boldface denotes optimal/greedy actions from state-action value. The use of variational information can significantly improve the power of the function factorization operators.}
\end{table}

Table~1b-1f shows the learning results of selected algorithms, QTRAN and LSF-SAC can learn a policy that each agent jointly takes the optimal action conditioning only on their local observations, meaning successful factorization. DOP falls into the sub-optimum caused by miscoordination penalties, similar to VDN and QMIX, which are limited by additivity and monotonicity constraints. Although QTRAN managed to address such limitations with more general value decomposition, as pointed out in later works \cite{mahajan2019maven} that it poses computationally intractable constraints that can lead to poor empirical performance on complex MARL domains.  It is also worth noting that LSF-SAC can find the optimal joint action under the monotonic constraints by providing variational information, however, its joint action value estimation will still be restricted by such limitation; this indicates that the multi-agent entropy maximization design and the utilization of latent state information can significantly enhance the exploration policies and improve the power of the monotonic factorization operators in a mixing network like QMIX.

Besides the single-state matrix game example shown in Table 1, we can also consider a multi-state problem with two agents, A and B. Let $(o^{(A)}_1,o^{(B)}_1)$ and $(o^{(A)}_2,o^{(B)}_2)$ be the two agents' observations in two different states $s_1$ and $s_2$. Providing latent information $m_{B}$ conditioned on $o^{(B)}_1$ and $o^{(B)}_2$ will enable Agent A to better estimate its local utility $Q_{A}(o^{(A)},m_{B})$ in the two states $s_1$ and $s_2$. Thus, with the latent information $m_{A}$ and $m_{B}$, 
the joint action-value function estimate with a mixing network $f$ is given by $Q_{tot}=f(Q_{A}(o^{(A)},m_{B}), Q_{B}(o^{(B)},m_{A}))$, which is able to represent a larger class of functions than $Q_{tot}=f(Q_{A}(o^{(A)}), Q_{B}(o^{(B)}))$, for the goal of estimating $Q^*(o^{(A)},o^{(B)})$.

\begin{figure}[ht]
\centering
\includegraphics[width=0.34\textwidth]{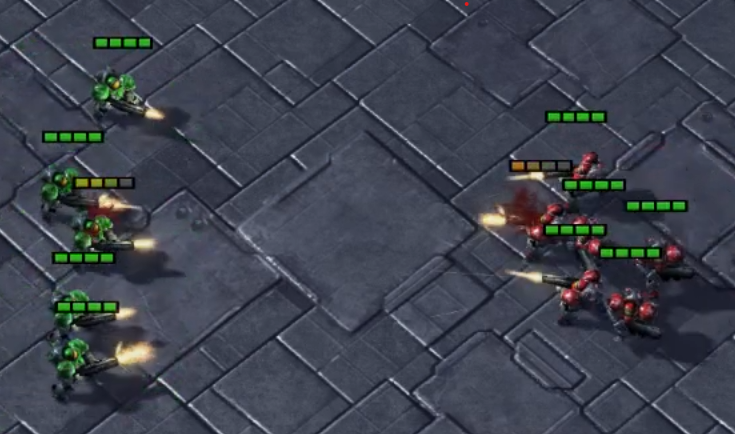}
\caption{An illustration of SMAC benchmark on map \texttt{5m\_vs\_6m}, where the testing algorithm is to control the 5 marines on the left (marked green), combating with 6 marines controlled by the game built-in AI on the right (marked red).}
\end{figure}
 
\subsection{Predator-Prey Environments}

We first evaluate the performance of our baseline algorithms on a partially-observable multi-agent environment Predator-Prey environment as described in \cite{bohmer2020deep}. This environment involves 8 predators cooperating to catch 8 AI-controlled prey units on a 10x10 grid, with successful captures requiring at least two predators to surround and capture a prey unit simultaneously. Our aim is to test the algorithms' ability to handle relative over-generalization and monotonicity constraints. More details are provided in the Appendix on this environment. In this relatively easy testing environment, we observe satisfying final results compared to SOTA works. Although, at the beginning of the training, a larger shaded area indicates a more volatile training procedure, this could be due to the insufficient training of the information generation module at its earlier stage demonstrating the effect of the overhead from the information sharing mechanism.

\begin{figure}[ht]
\centering
\includegraphics[width=0.5\textwidth]{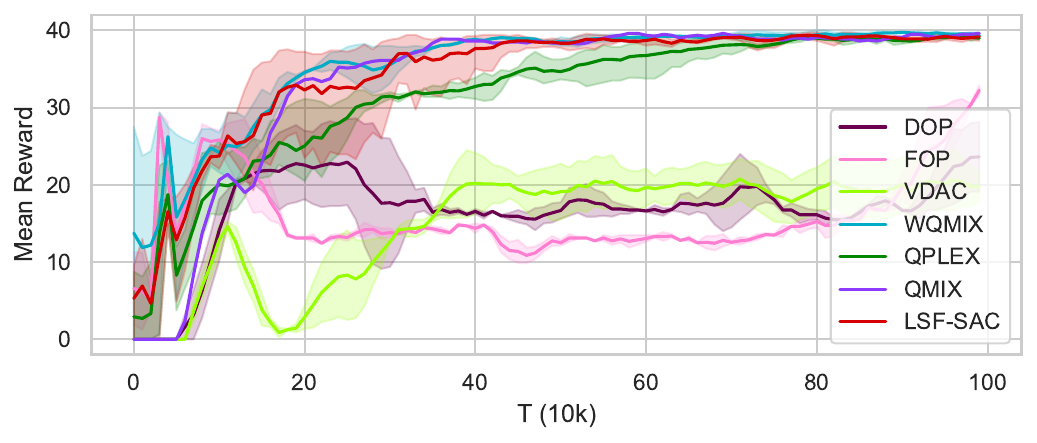}
\caption{ {Results on Predator-Prey Environments}}
\end{figure}

\subsection{Decentralised Starcraft II micromanagement benchmark}

To further assess the effectiveness of our approach, we benchmark its performance against various state-of-the-art multi-agent reinforcement learning (MARL) methods on selected scenarios from the StarCraft Multi-Agent Challenge (SMAC) \cite{samvelyan2019starcraft}. In Appendix A.3 we provide 

We then perform several ablation studies to analyze the factors that contribute to the performance. It is worth noting that the StarCraft Multi-Agent Challenges (SMAC) are affected by several code-level optimizations techniques, i.e., hyper-parameter tuning, as also found by \cite{hu2021riit}, some works are relying on heavy hyper-parameters tuning to achieve results that they otherwise cannot. Consistent with previous work, we carry out the test with the same hyper-parameters settings across all algorithms. More details about the algorithm implementation and settings can be found in Appendix {\bf C}.

\begin{figure*}[ht!]
	\centering
	\begin{subfigure}[t]{0.45\textwidth}
		\centering
		\includegraphics[width=\textwidth]{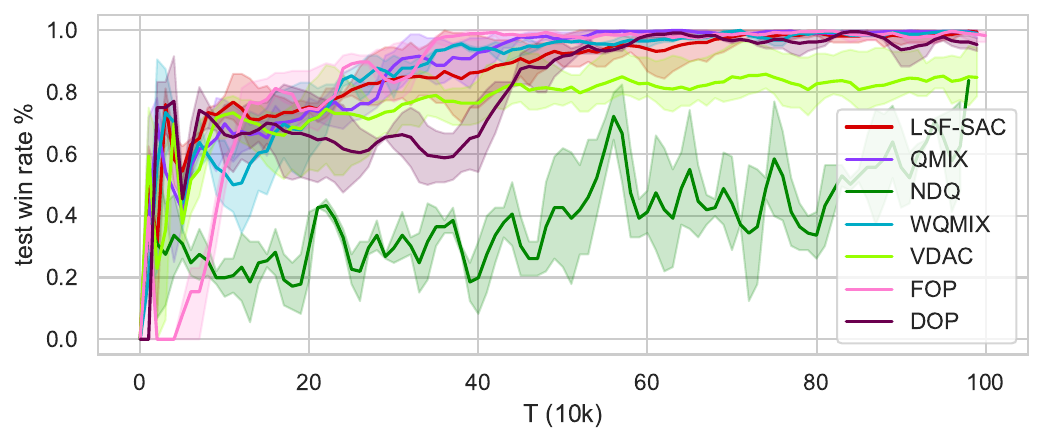}
		\caption{\small 3m (easy)}
		\label{fig:3m}
	\end{subfigure}
	\begin{subfigure}[t]{0.45\textwidth}
		\centering
		\includegraphics[width=\textwidth]{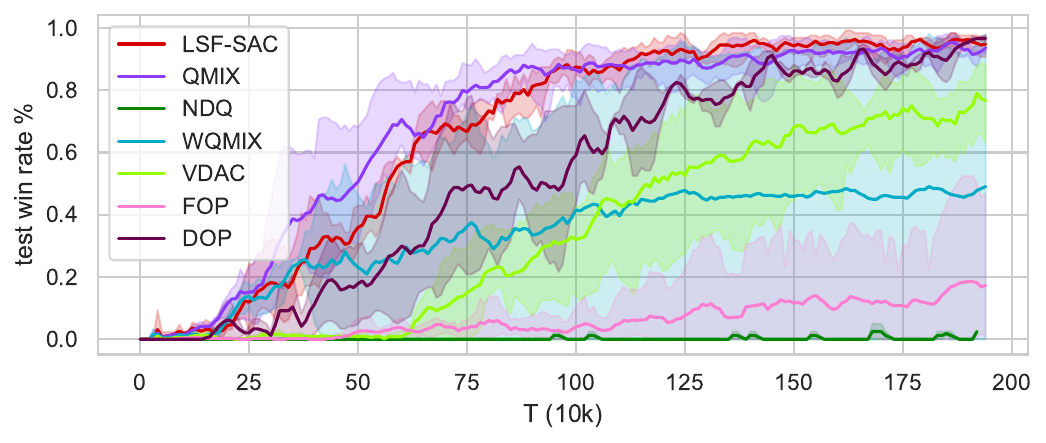}
		\caption{\small 3s5z (easy)}
		\label{fig:3s5z}
	\end{subfigure}

	\begin{subfigure}[t]{0.45\textwidth}
		\centering
		\includegraphics[width=\textwidth]{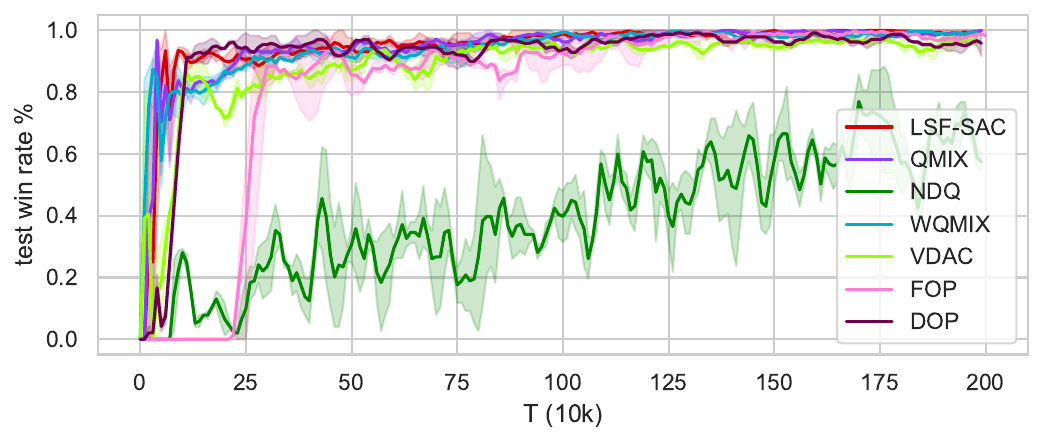}
		\caption{\small 8m (easy)}
		\label{fig:8m}
	\end{subfigure}
		\begin{subfigure}[t]{0.45\textwidth}
		\centering
		\includegraphics[width=\textwidth]{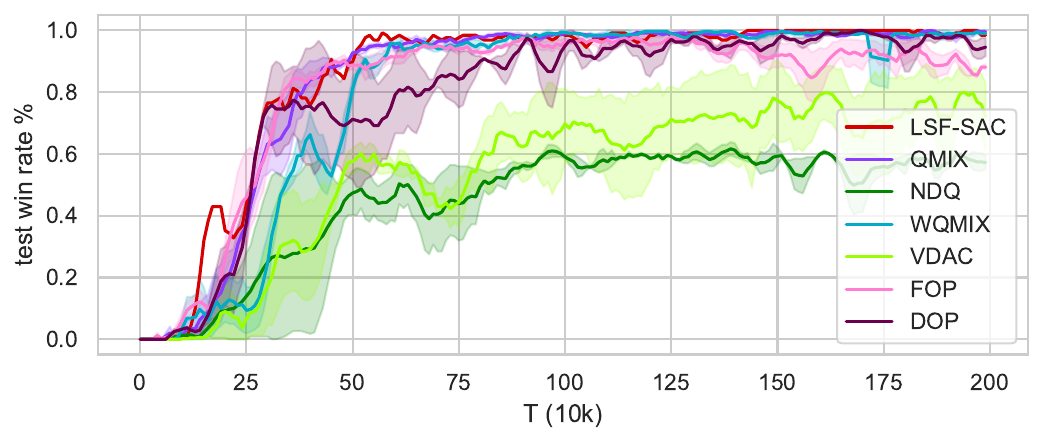}
		\caption{\small 1c3s5z (easy)}
		\label{fig:1c3s5z}
	\end{subfigure}

	\caption{{Results of 4 easy maps on the SMAC benchmark.}}
	\label{fig:smac}
\end{figure*}

\begin{figure*}[ht!]
	\begin{subfigure}[t]{0.325\textwidth}
		\centering
		\includegraphics[width=\textwidth]{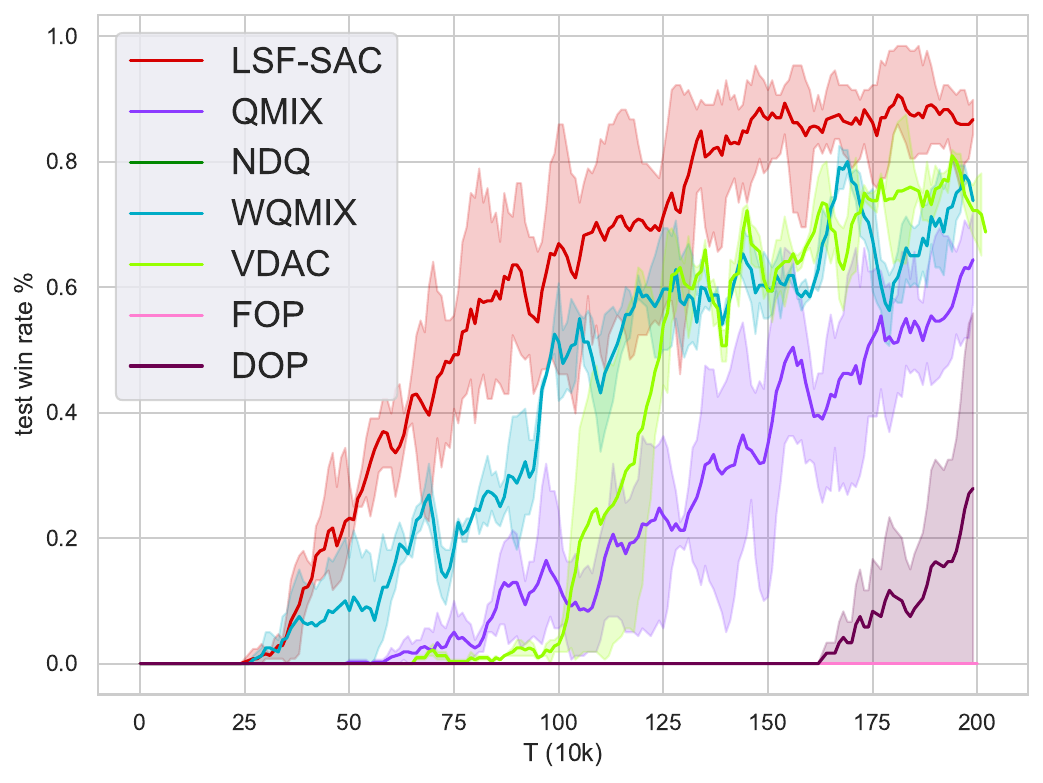}
		\caption{\small 27m\_vs\_30m (hard)}
		\label{fig:27m_vs_30m}
	\end{subfigure}
	\begin{subfigure}[t]{0.325\textwidth}
		\centering
		\includegraphics[width=\textwidth]{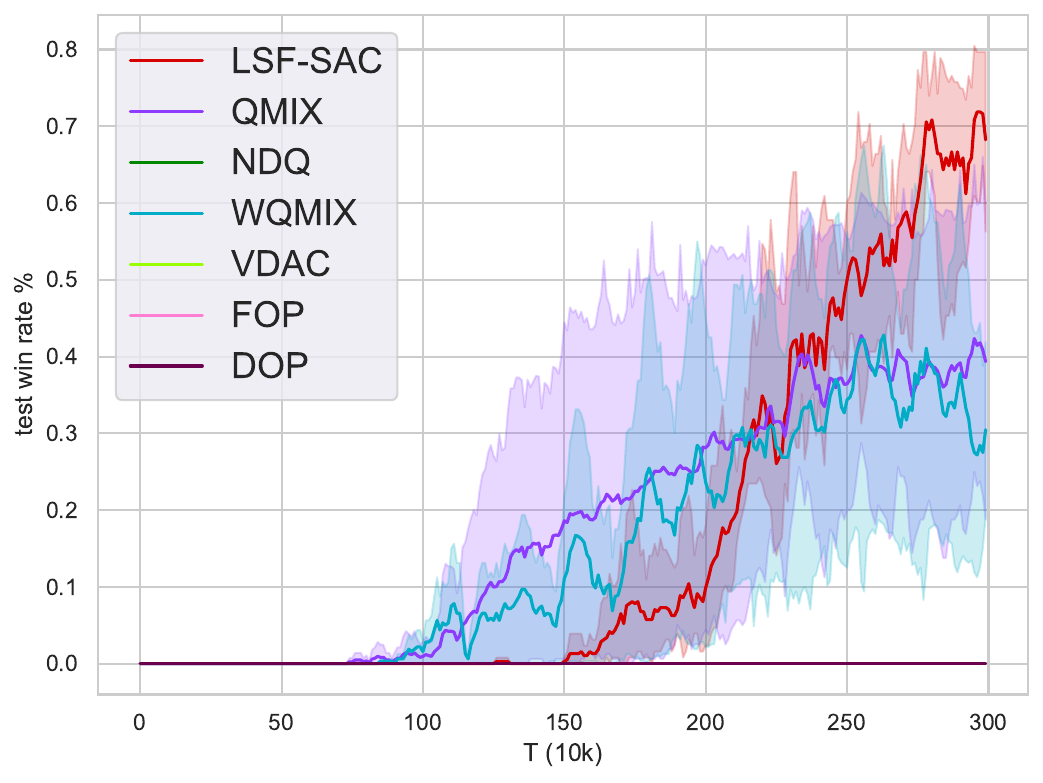}
		\caption{\small 3s\_vs\_5z (hard)}
		\label{fig:3svs5z}
	\end{subfigure}
	\begin{subfigure}[t]{0.325\textwidth}
		\centering
		\includegraphics[width=\textwidth]{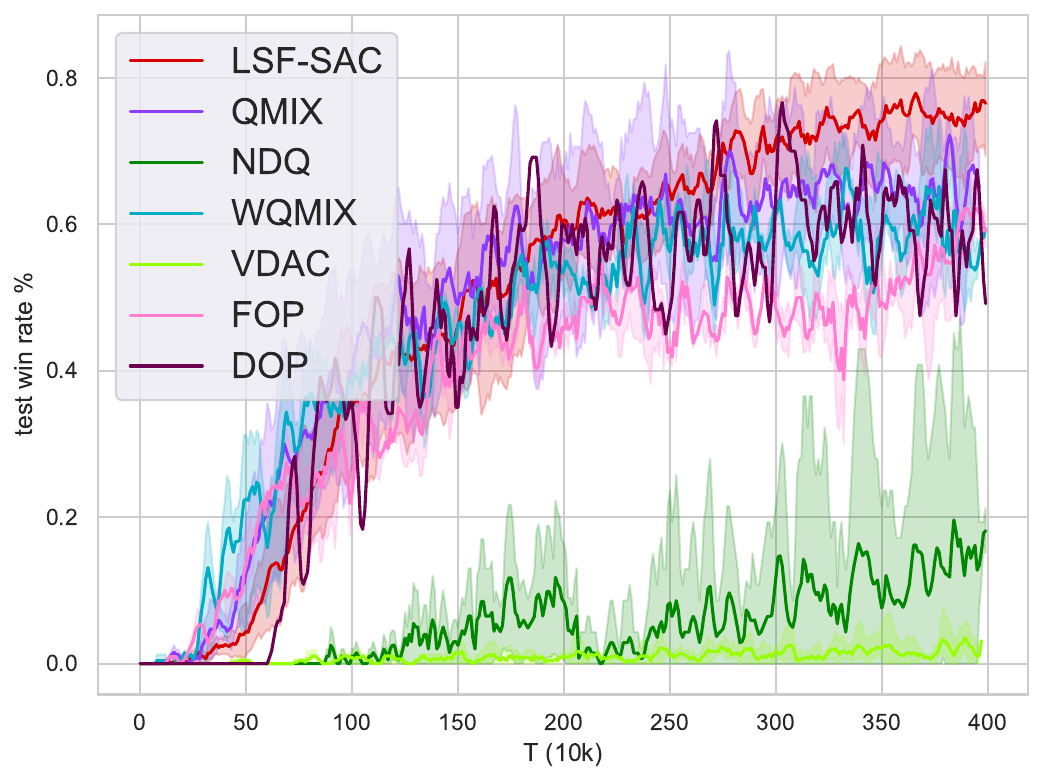}
		\caption{\small 5m\_vs\_6m (hard)}
		\label{fig:5mvs6m}
	\end{subfigure}
	
		\begin{subfigure}[t]{0.325\textwidth}
		\centering
		\includegraphics[width=\textwidth]{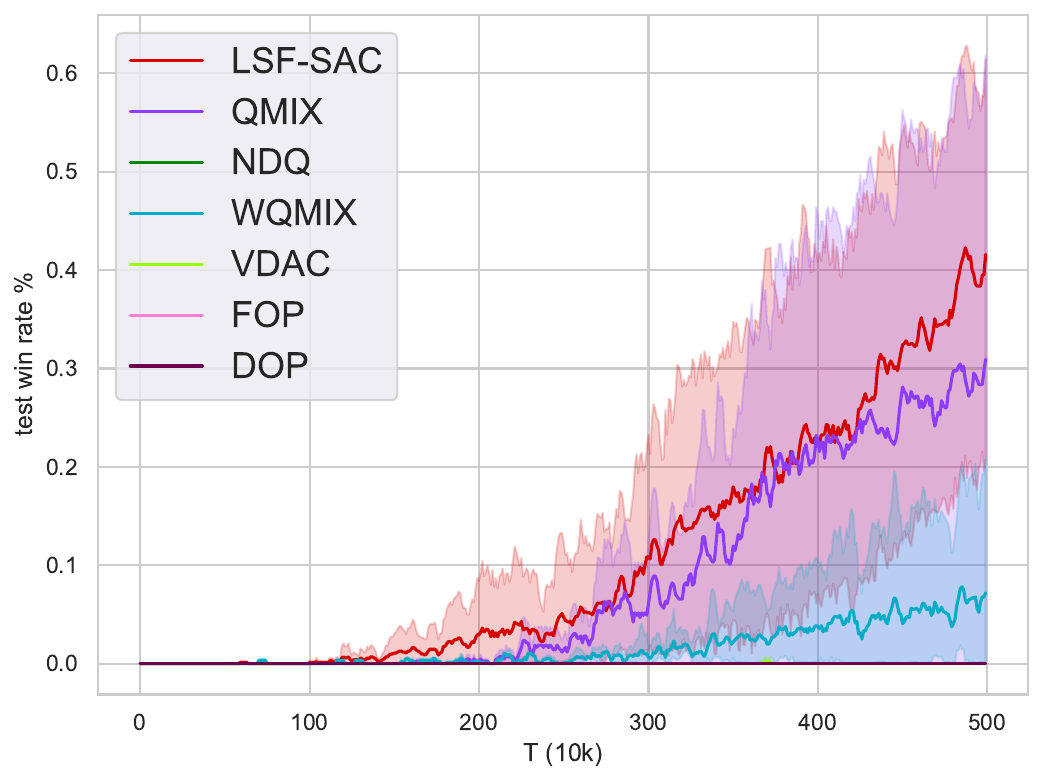}
		\caption{\small 6h\_vs\_8z (super hard)}
		\label{fig:6hvs8z}
	\end{subfigure}
		\begin{subfigure}[t]{0.325\textwidth}
		\centering
		\includegraphics[width=\textwidth]{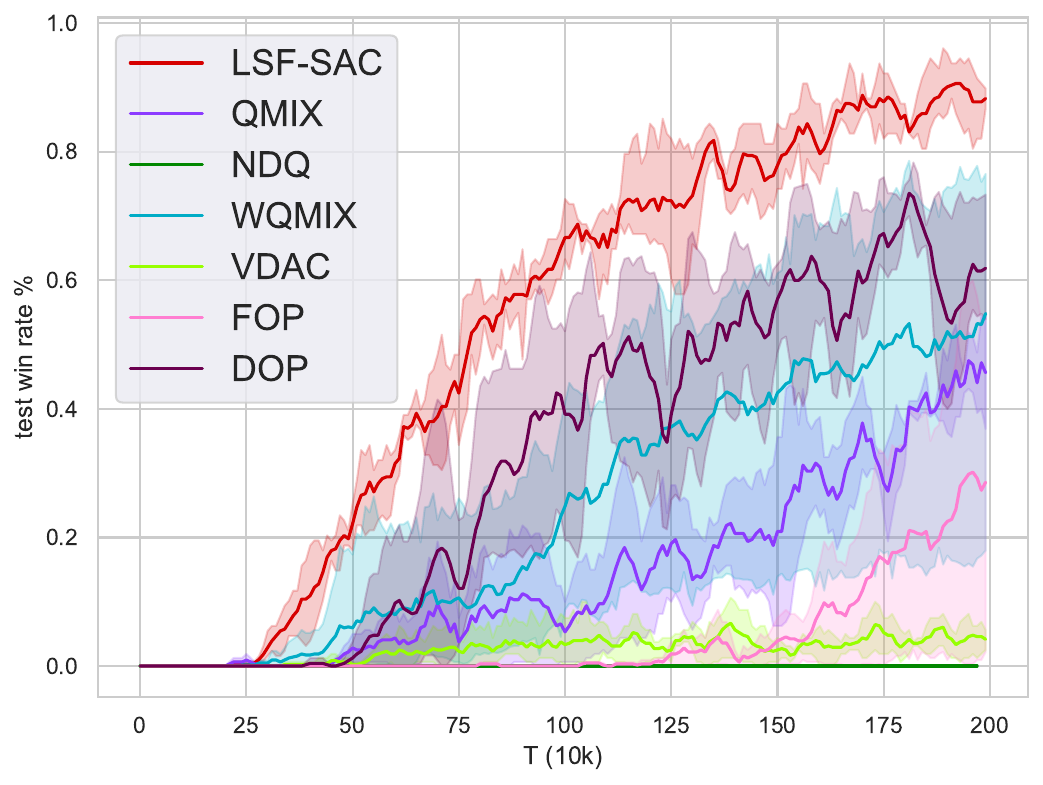}
		\caption{\small MMM2 (super hard)}
		\label{fig:mmm2}
	\end{subfigure}
	\begin{subfigure}[t]{0.325\textwidth}
		\centering
		\includegraphics[width=\textwidth]{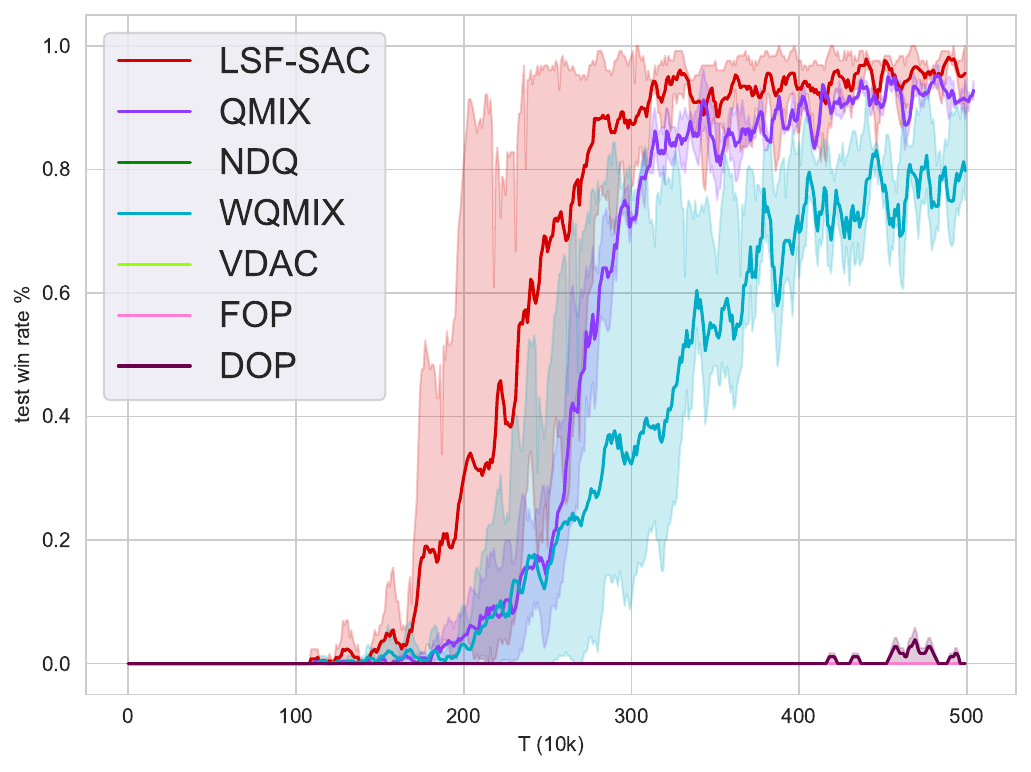}
		\caption{\small corridor (super hard)}
		\label{fig:corridor}
	\end{subfigure}
	\caption{{{{{Results of hard and super hard maps on the SMAC benchmark.}}}}}
	\label{fig:smac_hard}
\end{figure*}

In the SMAC benchmark\footnote{In this paper, all SMAC experiments are carried out utilizing the latest \texttt{SC2.4.10}, performance is always not comparable across versions. We implemented our algorithm based on an open-sourced codebase \cite{hu2021riit}.}, each agent is responsible for controlling a unit that collaborates with other friendly units in combat against the game's built-in AI-controlled units. The combat can take on a symmetric form, where both parties have access to the same units, or it can be asymmetric. Our testing is conducted on 10 different maps that cover all difficulty levels, including 4 easy maps (\texttt{3m, 3s5z, 8m, 1c3s5z}), 3 hard maps (\texttt{3s\_vs\_5z, 5m\_vs\_6m, 27m\_vs\_30m, }), and 3 super-hard maps (\texttt{6h\_vs\_8z, corridor, MMM2, 27m\_vs\_30m}). We selected these maps based on criteria such as the size of the action space (\texttt{27m-vs-30m}), the need for advanced exploration strategies (corridor), and the requirement for a high level of coordination between agents (\texttt{6h\_vs\_8z}). The same default environment setting was used for all benchmark algorithms in our testing, and each baseline algorithm was trained using 4 random seeds and evaluated every 10,000 training steps with 32 testing episodes. Further details on the environment setup and hyperparameter settings can be found in Appendix A.3.3. We compare LSF-SAC with several state-of-the-art MARL algorithms as baselines. We choose two decomposed actor critic methods: FOP \cite{zhang2021fop} and DOP \cite{wang2020dop}, one decomposed policy gradient method: VDAC \cite{su2021value}, three decomposed value based method: WQMIX \cite{rashid2020weighted}, QPLEX \cite{wang2020qplex} and QMIX \cite{rashid2018qmix}\footnote{In this section we refer WQMIX to ow-qmix as it shows a generally better performance than cw-qmix.}, and finally a communication based value-based method: NDQ \cite{wang2019learning}.

\subsection{General Results}

Following the practice of previous works \cite{samvelyan2019starcraft}, for every map result, we compare the winning rate and plot the median with the shaded area representing the highest and lowest range from testing results in Figure 2. In general, we observe LSF-SAC achieves strong performance on all selected SMAC maps, notably it outperforms the state-of-the-art algorithms or achieves faster and more stable convergence at a higher win rate. Note that LSF-SAC performs exceptionally well on testing maps with challenging tasks that require more state information or substantial cooperation.  Previous research has shown that there exists a performance gap between state-of-the-art (SOTA) value-based methods and policy gradient methods, particularly on maps that require the use of extensive exploration techniques.

In easy scenarios, almost all algorithms perform well. As the built-in AI would tend to attack the nearest enemy, by pulling back the friendly unit with a lower health value is a simple strategy to learn for winning. No significant performance gap was observed except for the training converging speed.

Within hard maps, LSF-SAC is able to train a usable policy that outperforms all baseline algorithms. On \texttt{27m\_vs\_30m} and \texttt{MMM2}, LSF-SAC performs exceptionally in terms of the convergence speed and the final performance. On \texttt{corridor}, LSF-SAC and the selected two value-based methods are able to learn a model, with our method converging faster with slightly better performance, while policy-based methods suffer from this map as it requires more exploration to find the specific trick in winning this challenging scenario. 
On \texttt{5m\_vs\_6m}, although within a similar performance range, LSF-SAC converges to a policy with lower variance and slightly better performance in the end. Finally, on \texttt{6h\_vs\_8z}, which is a super hard map that requires extensive exploration techniques, LSF-SAC achieves both faster convergence and better performance by a large margin as compared to the selected baselines.

It is also worth noting that the performance gap between value-based and policy-based methods  still exists even for the state-of-the-art methods, while LSF-SAC as a policy-based method not only narrows such gap but also achieves remarkable performance.

\subsection{Ablation study}

In this section, we perform a comparison between LSF-SAC and several modified algorithms to understand the contribution of different modules in LSF-SAC.
We choose one of the previously tested SMAC maps: \texttt{MMM2}. Each experiment is repeated with three independent runs with random seeds with their median results presented.

\subsubsection{Ablation 1} First, we consider the setting of LSF-SAC without the extra state info encoding (Purple part in Fig.1) as MASAC. This demonstrates how multi-agent soft-actor-critic works alone. It
highlights the importance of latent state information by comparing the results of MASAC against the original LSF-SAC.

\subsubsection{Ablation 2}  We also consider a fixed temperature design as LSF-SAC\_Fixed\_$\alpha$ with fixed $\alpha = 1.0$ (MASAC $\alpha=1.0$); this is to understand the effectiveness of the design in automatically updating the temperature $\alpha$.

\subsubsection{Ablation 3} We then consider the implementation of a multi-agent soft-actor-critic with value decomposition as MASAC, and the implementation of multi-agent advantage actor-critic with value decomposition as MAA2C, which can be considered as QMIX under a SAC and A2C setting, respectively \cite{su2021value}. This is to find the contribution of soft-actor-critic in enhancing exploration.

\subsubsection{Ablation 4} Finally we note that the original (single-agent) soft-actor-critic algorithm \cite{haarnoja2018soft} and several other works use two independently trained soft Q-functions and use the minimum of the two as the policy for optimizing, as \cite{hasselt2010double,fujimoto2018addressing} points out that policy steps are known to degrade the performance of value-based methods, e.g. in \cite{pu2021decomposed} they train with $\mathcal{L}(\theta)=[(r_{t}+\gamma\min _{j \in 1,2} Q_{tot}((\boldsymbol{s}_{t}^{'}, \boldsymbol{\tau}_{t}^{'}, \boldsymbol{a}_{t}^{'};\theta^{-}_{j})))-Q_{tot}(\boldsymbol{s}_{t}, \boldsymbol{\tau}_{t}, \boldsymbol{a}_{t};\theta))^{2}]$. Their performance comparison can be found in the ablation studies as MASAC\_DoubleQ \cite{pu2021decomposed}. This is to find if TD advantage with double Q learning is more stable under MARL when combined with value function decomposition. 

\begin{figure}[ht]
\centering
\includegraphics[width=0.5\textwidth]{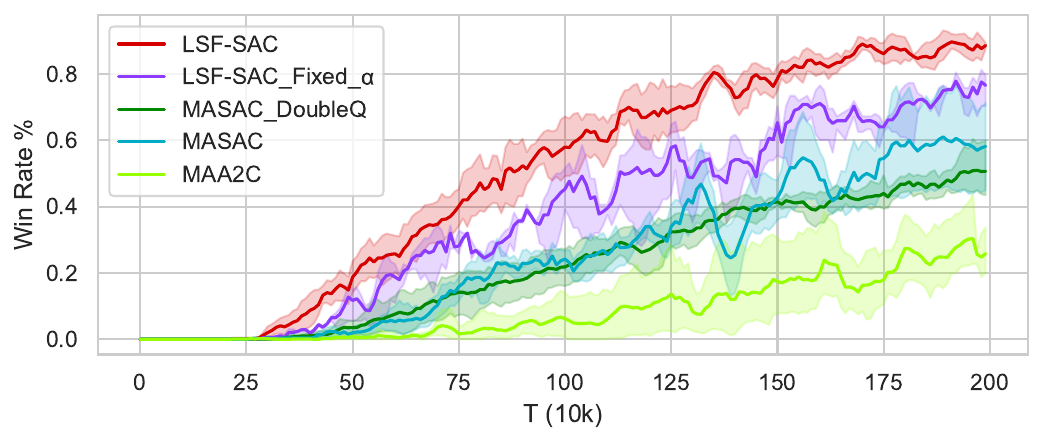}
\caption{ {{Ablation Results on MMM2}}}
\end{figure}
\textcolor{white}{Ablation} 

\subsection{Ablation Results}
By comparing the results of MASAC and LSF-SAC, we observe an improvement in both maps regarding the performance of LSF-SAC, which confirms the contribution of the latent state information assisted value decomposition design. 

Also,  LSF-SAC with $\alpha$=1.0 is able to achieve a higher winning rate and faster convergence than MASAC. The performance gap between LSF-SAC and MASAC demonstrates the importance of the proposed latent assistive information and our design of entropy maximization specialized for value decomposition methods. The performance gap between LSF-SAC and LSF-SAC with fixed $\alpha$ indicates the necessity of self-updating temperature term in balancing the trade-off between promoting exploration and maximizing the expected rewards.

Finally, although MSAC\_DoubleQ delivers a learnable policy at a plodding pace, this could potentially be the result of a complex model and relatively continuous reward on this specific environment. Also, due to its redundant network size, we find that MSAC\_DoubleQ, with its double value function design, takes a significantly longer time for training. This proves TD advantage with a single value function might be sufficient to optimize multi-agent actor critics within value decomposition methods. Nevertheless, we observe the design of the double Q network demonstrated the most stable training process with the lowest variance among all ablated baselines.

\section{Conclusions}
In this paper, we propose LSF-SAC, a novel framework that combines latent state information assisted individual value estimation for joint value function factorization and multi-agent entropy maximization, for collaborative multi-agent reinforcement learning under the CTDE paradigm. We introduce an information-theoretical regularization method for optimizing the latent state information assisted latent information generator to efficiently and effectively utilize extra state information in individual value estimation, while CTDE can still be maintained through a soft-actor-critic design. We also propose one possible implementation of expanding the off-policy maximum entropy deep reinforcement learning to the multi-agent domain with latent state information. 

Empirical results  show that our framework significantly outperforms the baseline methods on the SMAC environment. We further analyze the key factors contributing to the performance in our framework by a set of ablation studies. In future works, we plan to focus on expanding the proposed method with better generation and utilization of the extra state information with theoretical demonstrations of its assisting benefits.

\ifCLASSOPTIONcaptionsoff
  \newpage
\fi



%

\end{document}